# Thermally activated processes of the phase composition and structure formation of the nanoscaled Co–Sb films

R. A. Shkarban[1], Ya. S. Peresunko[1], E. P. Pavlova[1], S. I. Sidorenko[1], A. Csik[2], Yu. N. Makogon[1]

[1] National Technical University of Ukraine "Kyiv Polytechnic Institute," Kiev, Ukraine
[2] Institute of Nuclear Research of the Hungarian Academy of Sciences (ATOMKI), Debrecen, Hungary

It is investigated the formation of the phase composition and structure in the nanoscaled $CoSb_x$ (30 nm) ($1.82 \leq x \leq 4.16$) films deposited by the method of molecular-beam epitaxy on the substrates of the oxidated monocrystalline silicon at 200°C and following thermal treatment in vacuum in temperature range of 300–700°C. It is established that the films after the deposition are polycrystalline without texture. With increase in Sb content the formation of the phase composition in the films takes place in such sequence as this is provided by phase diagram for the bulky state of the Co–Sb system. At annealing in vacuum at temperature above 450–500°C a sublimation not only of the crystalline Sb phase but from the antimonides occurs. This is reflected on the phase composition change by following chemical reactions:

$CoSb_2 \rightarrow (600°C)$ $Sb\uparrow = CoSb$, $CoSb_3 \rightarrow (600°C)$ $Sb\uparrow = CoSb_2, CoSb_3 + Sb\uparrow \rightarrow (600°C)$ $CoSb_3$

and leads to increase in amount of the CoSb and $CoSb_2$ phases and decrease in amount of the $CoSb_3$. $CoSb_x$ (30 nm) ($1.8 < x < 4.16$) films under investigation are thermostable up to ~350°C.

## INTRODUCTION

Thermoelectricity, which is a priority in science and technology development, is based on the direct conversion of heat energy into electricity and vice versa. Skutterudite CoSb3 has a potential for use as a thermoelectric material [1–3].

Currently utilized thermo-electric materials have a maximum of efficiency $ZT$ only in the range of 1 [4–5]. $ZT$ is calculated by the formula $ZT = S^2 aT/(k_{el} + k_L)$, where $S$ is the Seebeck coefficient, $a$ is the electrical conductivity, $T$ is the absolute temperature, $k_{el}$ is the thermal conductivity of electrons, $k_L$ is the lattice thermal conductivity [2, 6]. According to theoretical calculation, $ZT$ increases with decrease in size and in nanomaterials, one can reach values >2, because of a decrease in lattice thermal conductivity [7].

The purpose of this study is to investigate the influence of deposition and thermal treatment conditions on the formation of the phase composition and structure in the nanoscaled $CoSb_x$ (30 nm) ($1.82 \leq x \leq 4.16$) films produced on oxidized monocrystalline silicon.

## EXPERIMENTAL PROCEDURE

30 nm thick $CoSb_x$ ($1.82 \leq x \leq 4.16$) films were produced by molecular-beam epitaxy on a substrate of thermally oxidized (100 nm thick $SiO_2$) monocrystalline Si (001). Antimony was deposited at a constant rate of 0.3 A/sec by an effuser heated to 470°C. Simultaneously, Co was co-deposited by electron-beam method. Phase composition modification was accomplished by varying the Co deposition rate from 0.027–0.049 A/sec. The pressure in the working chamber was $9.3 \cdot 10^{-11}$ Pa. The substrate temperature was 200°C. The Co content was determined by the luminous flux in the molecular beam. The Co deposition rate was measured by an EIES (Electron-Induced Emission Spectroscopy) optical system and was controlled by the SENTINEL III Leybold system during deposition. The deposition process was thus regulated in order to maintain the film thickness of 30 nm.

The film composition was determined by Rutherford backscattering (RBS) with an accuracy of ±1 at.% using $He^+$-ions at an acceleration energy of 1.7 MeV.

The film thickness was determined by simulation of the RBS-spectra using SIMNRA software for processing RBS-data. The statistical accuracy for the measurement of film thickness was ±1 nm. This was confirmed by X-ray reflectometry. The samples were annealed in vacuum and under nitrogen at 300–700°C for 0.5–5 h. The chemical composition of films was determined by Rutherford backscattering spectrometry. The phase composition and structure were characterized by X-ray diffraction (XRD) (Debye–Sherrer photo method with photographic registration of X-ray beams and using a diffractometer ULTIMA IV Rigaku with using Cu-$K_\alpha$-radiation and Bragg–Brentano geometry). The XRD data were corrected by the Rachinger algorithm [8]. The conductive properties of the films were investigated by resistometry using the four-point method. The quantitative change of the phase composition in the films was determined metallographically by the intercept method on the photographs of the surface obtained using a scanning electron microscope. The relative error of this method is ~4%.

## RESULTS AND DISCUSSION

Figure 1 shows the XRD patterns and photographs of the XRD patterns of the $CoSb_x$ ($1.82 \leq x \leq 4.16$) films after deposition. An identification of the phase composition revealed that, in the $CoSb_{1.82}$ (64.5 at.% Sb) film, the antimonide of $CoSb_2$ was formed with a monoclinic crystal lattice (Fig. 1*a*). According to the photographs of the XRD patterns, the films under investigation were polycrystalline without texture (Fig. 1*b*). Also, an increased Sb content resulted in the formation of polycrystalline skutterudite phase of $CoSb_3$ with a cubic lattice. This two-phase state is preserved in the films with up to 74.6 at.% Sb content. In this Sb content interval, the intensity ratio of the diffraction maxima of $I(210)CoSb_2/I(310)CoSb_3$ decreased, as is evidenced by the increase in $CoSb_3$ phase amount and decrease in $CoSb_2$ with increased Sb content (Fig. 2).

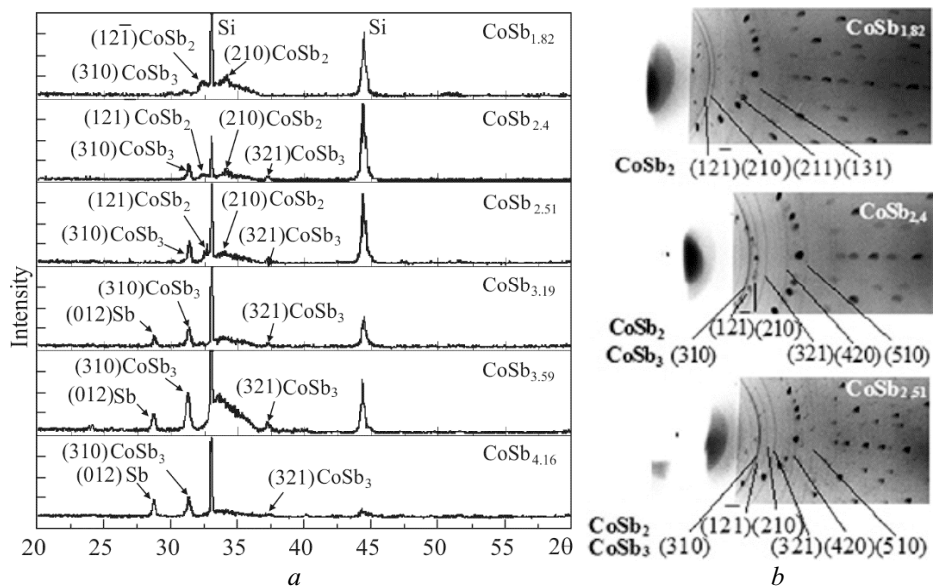

*Fig. 1.* XRD patterns (*a*) and photographs of XRD patterns (*b*) of the $CoSb_x$ ($1.82 \leq x \leq 4.16$) films after deposition

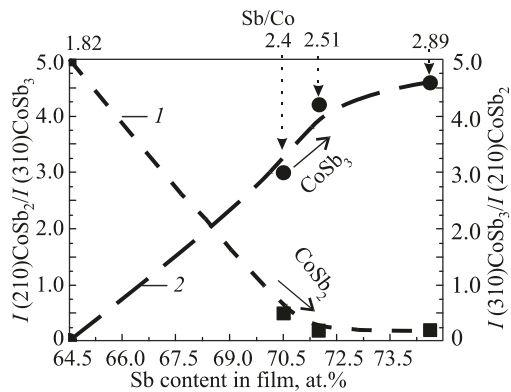 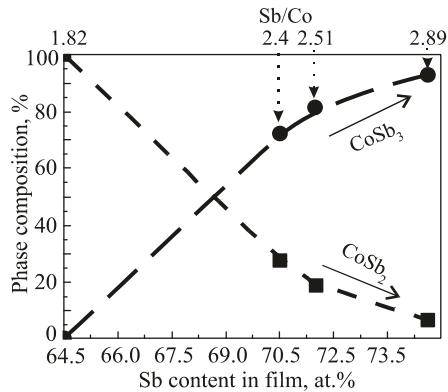

*Fig. 2*. The XRD integral intensity ratio of $I(210)CoSb_2/I(310)CoSb_3$ (*1*) and $I(310)CoSb_3/$
$/ I(210)CoSb_2$ (*2*) for the as-deposited $CoSb_x$ ($1.82 \leq x \leq 2.89$) films

*Fig. 3*. Change of phase composition in the as-deposited $CoSb_x$ ($1.82 \leq x \leq 2.89$) films

Figure 3 shows the change in the film phase composition with increased Sb content in the interval from 64.5 to 74.6 at.%, according to the results of the quantitative metallographic analysis of the SEM-images. The single-phase structure is observed in both $CoSb_{1.82}$ and $CoSb_{2.89}$ films. In the $CoSb_{1.82}$ film, the $CoSb_2$ phase is formed and in the $CoSb_{2.89}$ film, the $CoSb_3$ phase is fixed. At intermediate compositions, these phases coexist. Meanwhile, with increased Sb content in films, the skutterudite $CoSb_3$ amount increases and the $CoSb_2$ phase amount decreases. The use of these two methods demonstrated good correlation with the change of the phase composition in as-deposited films, subject to Sb content (Figs. 1 and 3).

In the as-deposited $CoSb_x$ ($3.19 \leq x \leq 4.16$) films with higher Sb content, the two-phase composition was also observed. In addition to $CoSb_3$ in the films, the crystalline phase of Sb (Fig. 1) was also formed. The change of the ratio of diffraction peak intensities of $I(012)Sb/I(310)CoSb_3$ demonstrates that increasing Sb content from 76.1 to 80.6 at.% Results in an increase of the amount of the crystalline Sb phase (Fig. 4).

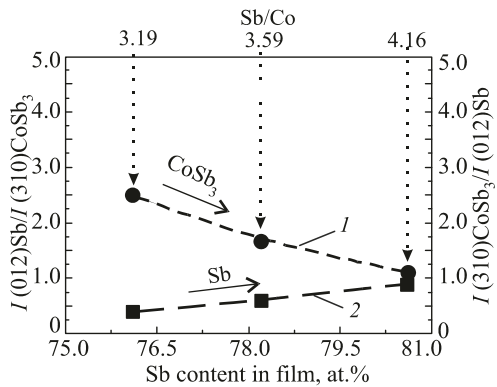

*Fig. 4*. The XRD integral intensity ratio of $I(012)Sb/I(310)CoSb_3$ (*1*) and $I(310)CoSb_3/I(012)Sb$ (*2*)
for the as-deposited $CoSb_x$ ($3.19 \leq x \leq 4.16$) films

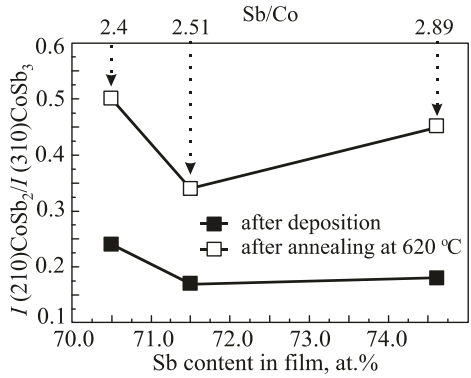

*Fig. 5.* The XRD integral intensity ratio of *I*(210)CoSb$_2$/*I*(310)CoSb$_3$ for the CoSb$_x$ (2.4 ≤ *x* ≤ 2.89) films after deposition and annealing in vacuum at 620°C for 30 sec

TABLE 1. Phase composition of the As-deposited CoSb$_x$ (1.82 ≤ *x* ≤ 4.16) Films

| Substrate temperature, °C | Sb content in the film, at.% | | | |
|---|---|---|---|---|
| | 64.5 | 64.5–70 | 70–74.6 | 76.1–80.6 |
| | Ratio Sb/Co in the film | | | |
| | 1.82 | 1.82–2.3 | 2.3–2.89 | 3–4.16 |
| 200 | CoSb$_2$ | CoSb$_2$ + CoSb$_3$ | CoSb$_3$ + (CoSb$_2$) | CoSb$_3$ + Sb |

Table 1 demonstrates the results of the XRD structure and phase analysis of the films under investigation. It should be noted that in nanoscale films deposited at the substrat temperature of 200°C, a good coincidence was observed in the phase composition with the phase diagram for bulk state.

Annealing of CoSb$_x$ (2.4 ≤ *x* ≤ 2.89) films in vacuum causes a change in their phase composition.

After annealing at 620°C, the ratio *I*(210)CoSb$_2$/*I*(310)CoSb$_3$ increases with a noted absence of texture (Fig. 5). This indicated an increase in the CoSb$_2$ amount.

The as-deposited CoSb$_x$ (3.19 ≤ *x* ≤ 4.16) films had a two-phase crystalline structure of a skutterudite phase of CoSb$_3$ and a crystalline Sb phase. The change in the ratio of the diffraction peak intensities of *I*(210)CoSb$_2$/*I*(310)CoSb$_3$ after annealing in the Sb-enriched films of CoSb$_{3.59}$ and CoSb$_{4.16}$ demonstrates that the annealing under 500°C does not result in the phase changes (Fig. 6).

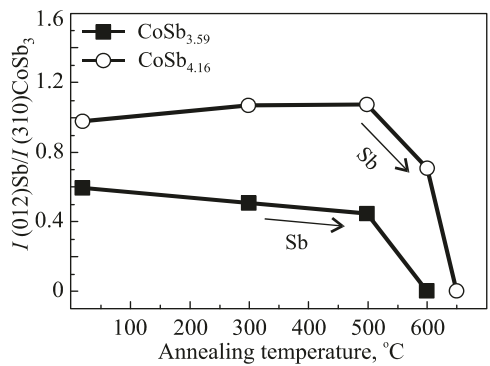

*Fig. 6.* XRD integral intensity ratio of *I*(012)Sb/*I*(310)CoSb$_3$ for the CoSb$_{3.59}$ and CoSb$_{4.16}$ films after annealing in vacuum at different temperatures

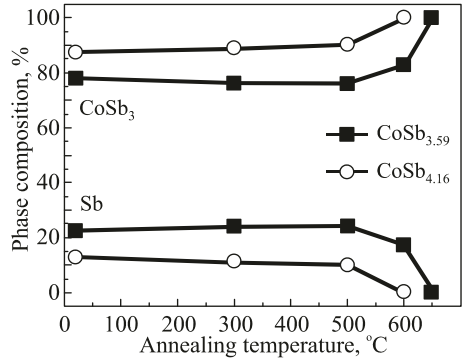

*Fig. 7*. Phase composition of the $CoSb_{3.59}$ and $CoSb_{4.16}$ films after annealing in vacuum at different temperatures for 30 sec

During annealing at higher temperatures, an intense sublimation of Sb occurred. After annealing at 600°C, the reflexes of the crystalline Sb were not observed and only the $CoSb_3$ phase remained.

Figure 7 shows the data of the quantitative analysis of the phase composition change in the films with the two-phase composition of ($CoSb_3$ + Sb) after annealing, determined using the ratio intensities of the diffraction lines for (012)Sb and (310)$CoSb_3$, according to a previously reported method [9]. The process of Sb sublimation from antimonides also holds for the annealing of bulk materials [10].

The process of Sb sublimation also occurs during annealing under nitrogen. According to the XRD data on the structure and phase analysis in the as-deposited $CoSb_{1.82}$ film with the lowest Sb content (64.5 at.% Sb), only the $CoSb_2$ phase was observed, and after annealing at 600°C, a two-phase state of $CoSb_2$ and Sb was formed. Before and after such annealing, the two-phase composition was preserved for the antimonides of $CoSb_2$ and $CoSb_3$ in the $CoSb_{2.4}$ and $CoSb_{2.51}$ films with Sb content of 70.5 and 71.5 at.%, respectively.

Figure 8 shows the data of the quantitative metallographic analysis of the SEM-images of the $CoSb_x$ ($1.82 \leq x \leq 2.51$) films after deposition and thermal treatment. After annealing, a quantitative change of the phase composition was observed. After annealing at 600°C, nearly 30% of the Sb phase appeared in the film with Sb content of 64.5 at.%. In films with 70.5 and 71.5 at.% Sb content, the amount of the $CoSb_2$ phase increased and the amount of the $CoSb_3$ skutterudite decreased.

This can be explained by a partial sublimation of Sb from the crystalline lattices of the CoSb and $CoSb_3$ antimonides during annealing both in nitrogen and in vacuum due to the following chemical reactions:

$CoSb_2 \xrightarrow{600°C} Sb\uparrow = CoSb_2 + CoSb$; $CoSb_3 \xrightarrow{600°C} Sb\uparrow = CoSb_3 + CoSb_2$. Thermal stability of the nanoscale skutterudite films of $CoSb_x$ ($3.19 \leq x \leq 4.16$) was preserved up to ~300–350°C (Fig. 9).

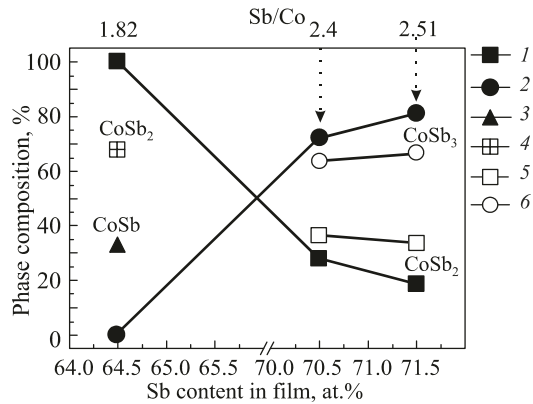

*Fig. 8*. Phase composition of the as-deposited (*1, 2*) films and the films post-annealed in nitrogen at 600°C for 30 sec (*3–6*)

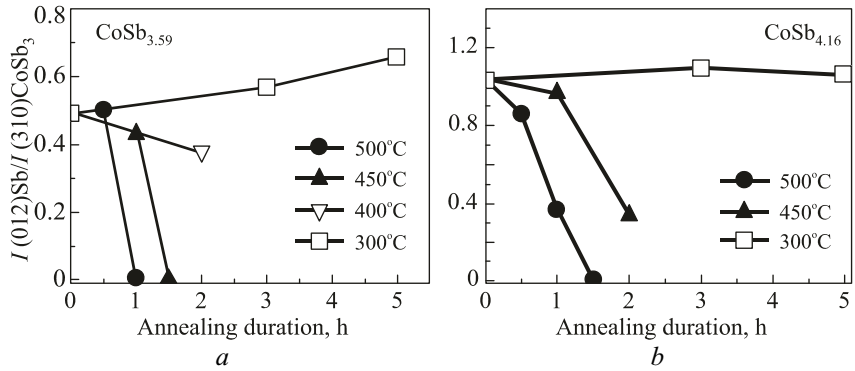

*Fig. 9.* Variation of the ratio of the diffraction peak intensities of $I(012)Sb/I(310)CoSb_3$ of the $CoSb_{3.59}$ (*a*) and $CoSb_{4.16}$ (*b*) films with annealing duration in vacuum at 300, 400, 450, and 500°C

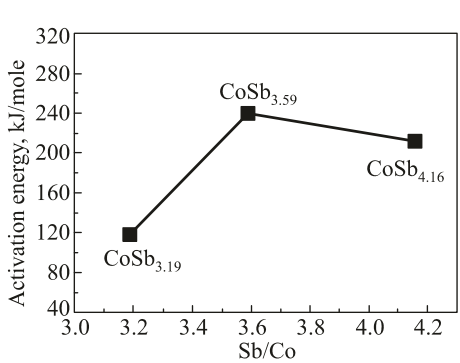

*Fig. 10.* Change of the activation energy of Sb sublimation in the $CoSb_x$ ($3.19 \leq x \leq 4.16$) films

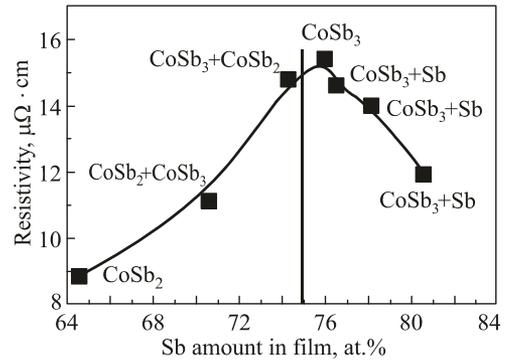

*Fig. 11.* Variation of the resistivity of $CoSb_x$ ($1.82 \leq x \leq 4.16$) with Sb amount

The activation energy for the Sb sublimation process was determined using the rate of the Sb sublimation at different annealing temperatures, according to the Arrhenius equality [10].

The sublimation process for crystalline Sb depends on the chemical composition of the films (Fig. 10).

The electrical-physical properties of the Co–Sb films depended on their phase composition. The dependence of the resistivity of the as-deposited films on the Sb amount has a parabolic character with a maximum of 15 μΩ · cm at the Sb concentration of 75 at.% (Fig. 11).

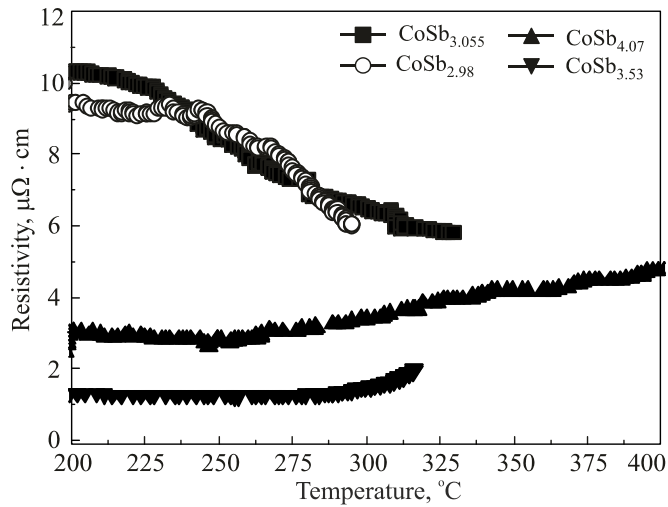

*Fig. 12.* Variation of the resistivity of the $CoSb_x$ ($2.98 \leq x \leq 4.07$) films with temperature

Skutterudite $CoSb_3$ is a semiconductor and has higher resistivity in comparison with the phases of $CoSb_2$ and Sb having a semiconductor and metallic type of the conductivity, respectively. The difference in the phase composition affects not only the absolute values of the film resistivity, but also their temperature dependence.

Therefore, for the $CoSb_{2.98}$ and $CoSb_{3.055}$ films, in which the $CoSb_3$ phase was generally present, the variation of the resistivity with the temperature has a semiconductor character (Fig. 12).

In the films with two-phase compositions from $CoSb_3$ and Sb, the temperature variation of the resistivity takes with a form typical for metals. According to the report [8], Sb has a metallic type of conductivity.

## CONCLUSION

It has been established that, during the deposition of $CoSb_x$ ($1.82 \leq x \leq 4.16$) films at the substrate temperature of 200°C, they form in a crystalline state. The films are in a polycrystalline state without texture. Good correlation between the film phase composition and the phase diagram for the bulk materials is observed. With increased Sb content, the formation of the phase composition occurs in the same sequence as is provided by the phase diagram for bulk materials of the Co–Sb system.

In $CoSb_x$ films, the phase formation sequence versus Sb amount is as follows:
- at 64.5 at.% Sb, the film formed is an antimonide of $CoSb_2$;
- from 64.5 to 75 at.% Sb, along with the $CoSb_2$ phase, the skutterudite of $CoSb_3$ is formed. $CoSb_3$ content increased with increased Sb;
- near 75 at.% Sb, the crystalline phase of $CoSb_3$ is formed;
- at greater than 75 at.% Sb, in addition to $CoSb_3$, the crystalline phase of Sb is formed.

With annealing in vacuum at temperatures higher than 450–500°C, Sb sublimation occurs, which is reflected in the change of the phase composition according to the following chemical reactions: $CoSb_2 \xrightarrow{600°C} Sb\uparrow = CoSb$, $CoSb_3 \xrightarrow{600°C} Sb\uparrow = CoSb_2$. This results in an increase in the CoSb and $CoSb_2$ amount and a decrease in the $CoSb_3$ amount.

The $CoSb_x$ ($1.82 \leq x \leq 4.16$) films are thermally stable up to ~350°C.

## ACKNOWLEDGMENTS

The authors would like to thank Prof. M. Albrecht, Dr. G. Beddies, PhD M. Daniel, and workers from Chemnitz University of Technology (Germany) for sample preparation, assistance in conduction of investigations and discussion of results.

This work was financially supported by the Deutsche Akademischer Austauschdienst (DAAD) in the frame of the Leonard-Euler-Program (Grant N 50744282).